\documentclass[conference]{IEEEtran}

\usepackage{hyperref}
\usepackage{caption}
\usepackage{adjustbox}
\usepackage{booktabs}
\usepackage{subcaption} 
\usepackage[font=footnotesize,skip=0pt]{caption}
\usepackage[shortlabels]{enumitem}
\usepackage{graphicx}
\usepackage{multicol}
\usepackage{float}
\usepackage{url}
\usepackage[table,xcdraw]{xcolor}
\usepackage{titlesec}
\titlespacing{\section}{0pt}{*1}{*0.5}
\usepackage{placeins}
\titleformat{\section}
  {\centering\bfseries\normalsize} % formatting: centered, bold, normal size
  {\thesection.}{1em}{}            % section number format
\usepackage{titlesec}

\titleformat{\subsection}[runin]
  {\bfseries\large}{\thesubsection}{1em}{}

\titleformat{\subsubsection}[runin]
  {\bfseries\normalsize}{\thesubsubsection}{1em}{}

\begin{document}

\title{DecoMind: A Generative AI System for Personalized Interior Design Layouts }
\author{Reema Alshehri, Rawan Alotaibi, Leen Almasri, Rawan Altaweel }
\maketitle

\vspace{0.5cm}

\begin{abstract}
This paper introduces a system for generating interior design layouts based on user inputs, such as room type, style, and furniture preferences. CLIP extracts relevant furniture from a dataset, and a layout that contains furniture and a prompt are fed to the Stable Diffusion with ControlNet to generate a design that incorporates the selected furniture. The design is then evaluated by classifiers to ensure alignment with the user’s inputs, offering an automated solution for realistic interior design.
\end{abstract}

\section{Introduction}
\IEEEPARstart{I}{nterior}  design has become increasingly popular as people seek more comfort and personalization in their living spaces. While hiring professional designers is common for full-home projects, redesigning a single room—such as a bedroom—may not justify the cost or effort involved in hiring such services.Additionally, many individuals who prefer to furnish their rooms using items from specific stores like IKEA often feel uncertain about whether suggested furniture—based on their selected categories (e.g., sofa, table)—will suit the room’s size, layout, and style.

 This lack of visual clarity often leads to hesitation and indecision.Both of these challenges—cost and uncertainty—are addressed by our proposed system, DecoMind. The name combines “Deco” (short for decoration) and “Mind,” reflecting the system’s goal of enabling smart, AI-assisted design decisions. By entering basic information such as room dimensions, type, preferred style, and furniture categories, users can visualize how recommended furniture—currently supporting only IKEA items—might appear in their space.

 DecoMind is intended primarily to assist individuals redesigning a single room or validating design choices before purchase, complementing rather than replacing professional interior designers.

\section{Related Work}
Several previous research papers have explored AI-driven interior design generators. 
The CLIP-Layout model \cite{cliplayout2023} introduced a method that learns visual correspondences between layout and text, enabling style-consistent and aesthetically pleasing scenes. However, this approach primarily focused on matching visual style and lacked structured control over furniture placement.

Another approach is DeepFurniture \cite{deepfurniture2020}, which uses object detection and re-ranking modules to identify furniture in room images and recommend compatible sets from a fixed dataset. Although useful for retrieval, it does not support generating new layouts or adapting to customized user constraints such as room size or door/window placement.

More recently, diffusion-based techniques have shown promise in generating realistic interior scenes. For example, CreativeDiffusion \cite{creativediffusion2025} integrates ControlNet with Stable Diffusion to guide image generation based on structural layouts. This model improves coherence between generated images and the room's physical structure, yet its focus is limited to visual realism and does not include user-specific furniture or interactive customization.

\vspace{0.2cm}
\noindent\textbf{Our Contribution:} Unlike previous works, our project \textit{DecoMind} offers a fully integrated system that:
\begin{itemize}
    \item Combines CLIP and Stable Diffusion for furniture-aware generation based on both spatial structure and user preferences.
    \item Allows users to define the room dimensions, door/window positions, furniture source (e.g., IKEA), and specific aesthetic styles.
    \item Automatically extracts suitable furniture, filters complex scenes, removes backgrounds, and generates structured layout images.
    \item Uses ControlNet-guided diffusion to synthesize realistic designs while preserving layout alignment.
    \item Classifies the generated output and evaluates it against user intent, providing feedback for future tuning.
\end{itemize}

This holistic approach bridges the gap between retrieval-based systems like DeepFurniture and style-guided models like CLIP-Layout, enabling a more personalized, interactive, and layout-consistent design experience.

\section{Data Description}

In developing the DecoMind system, we relied on three distinct datasets, given the lack of a unified dataset that encompasses room types, interior design styles, and the furniture associated with the store selected by the user for the generation process. As outlined in the introduction, the current version of our project is tailored exclusively to furniture available from the IKEA store. The datasets utilized are summarized  as
 follows:

\begin{itemize}
    \item {IKEA Furniture Dataset} \cite{ikeafurniture2025}: This dataset contains three folders: Train, Validation, and Test. Each folder includes images of IKEA furniture items (e.g., sofas, chairs, tables, etc.). It is primarily used to guide the model in generating images that feature IKEA furniture accurately.
    Additionally, we performed a custom filtering process to remove low-quality or irrelevant images, ensuring a cleaner and more accurate dataset for furniture representation.
    \item {House Rooms Image Dataset} \cite{houserooms2025}: This dataset is organized into multiple folders, with each folder representing a specific room type (e.g., bedrooms, kitchens, living rooms, etc.). It enables the model to recognize and differentiate between various types of rooms.

    \item {Interior Design Styles Dataset} \cite{interiordesignstyles2025}: This dataset consists of several folders, each corresponding to a particular interior style (e.g., modern, classic, minimalist, etc.). It helps the model identify and distinguish between various interior design styles.
    
    *Notes:The Interior Design Styles dataset was obtained from Kaggle, and it contains images originally sourced from Houzz.com. These images are subject to copyright restrictions. Their use is limited to academic and research purposes only, and redistribution is not allowed.

\end{itemize}

\section{Methodology}We have multiple stages in our study, each utilizing different methods for achieving the desired results.
The following diagram provides an overview of the entire system architecture, illustrating the main components and the flow between them:

\begin{figure}[H]
    \centering
    \includegraphics[width=0.45\textwidth]{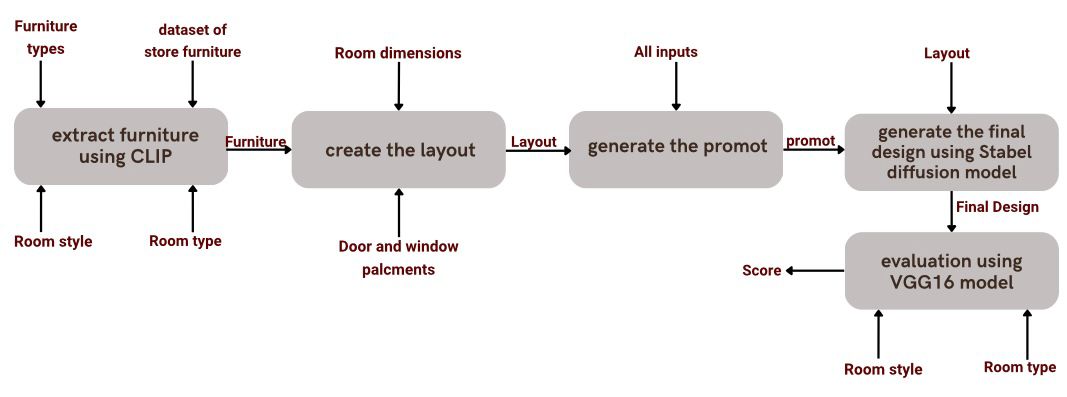} % Replace with actual file name
    \caption{Overview of the general workflow of the DecoMind system.}
    \label{fig:general_flow}
    \vspace{-1em}
\end{figure}

As shown in the diagram, the system is composed of several interconnected modules, starting from data preparation to image generation and furniture selection. Each component plays a specific role in ensuring the final design aligns with the user's input.A detailed explanation of each stage is provided below:

\begin{enumerate}[label=\Alph*.]  % a, b, c, ...
    \item \textbf{Furniture Extraction:}

The process begins with extracting furniture from the selected store. The extraction process is driven by user choices, which include room type, style, and furniture types. This is achieved using a trained CLIP model, which uses a dual-encoder architecture—one for processing images and the other for processing text. CLIP matches the provided text (room type, style, furniture types) with relevant images to ensure the best outputs.

\begin{figure}[H]
    \centering
    \includegraphics[width=0.3\textwidth]{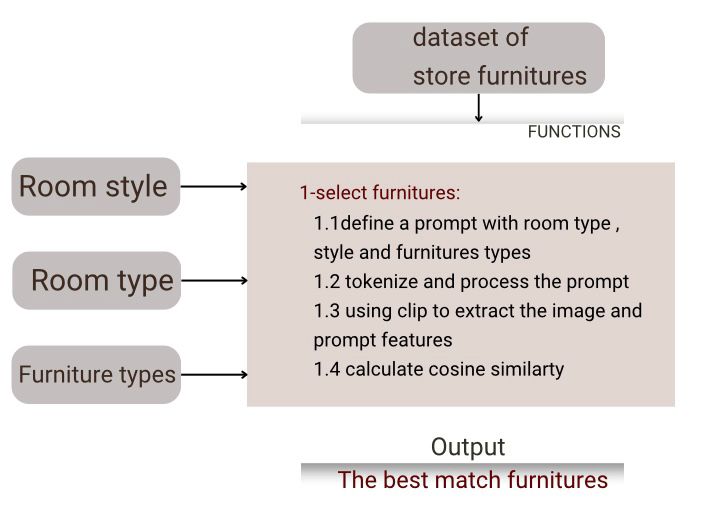}
    \caption{Extracting furniture.}
    \label{fig:fig1_furniture_extraction}
\end{figure}

    \item \textbf{Image Generation Using Stable Diffusion:}The next stage involves the main component of the system: the Stable Diffusion model. This model generates images from text inputs using a latent diffusion process. It has achieved impressive results in recent years but has limitations, particularly in using reference images. In our case, we wanted the model to generate interior designs with the furniture selected by the user. To overcome this limitation, we integrated ControlNet, which guides Stable Diffusion using additional input such as the layout of the room (room dimensions, window and door positions) and the extracted furniture from the previous stage.

-Stable Diffusion Architecture: Stable Diffusion uses a deep convolutional neural network with a transformer-based architecture for image generation. It operates by mapping the text prompt into a latent space, where it iteratively refines the image through a denoising process.

-ControlNet Architecture: ControlNet acts as a conditioning network, guiding the generative process by adding constraints from the layout and furniture data, ensuring that the generated design adheres to the required dimensions and object placements.

\begin{figure}[H]
    \centering
    \includegraphics[width=0.3\textwidth]{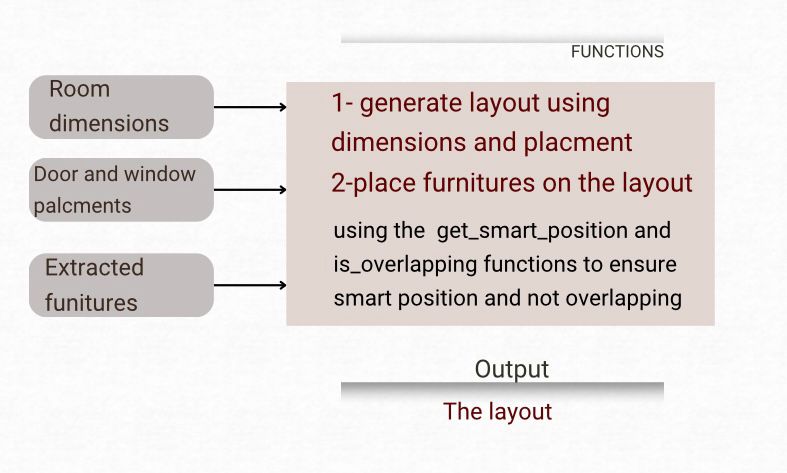}
    \caption{Generate the layout.}
    \label{fig:fig2_generate_layout}
\end{figure}

After extracting the furniture and generating the room layout, the subsequent step involved constructing a descriptive text prompt for the image generation model.During this process, we faced limitations related to the IKEA dataset (store dataset), as many furniture items lacked sufficient detail for effective generation, and the dataset did not include complete 3D views of the objects. To address these issues, we instructed the model in the prompt to “imagine and correct” the angles of the furniture
 objects and feel free to add more items to complete the design.

\begin{figure}[H]
    \centering
    \includegraphics[width=0.3\textwidth]{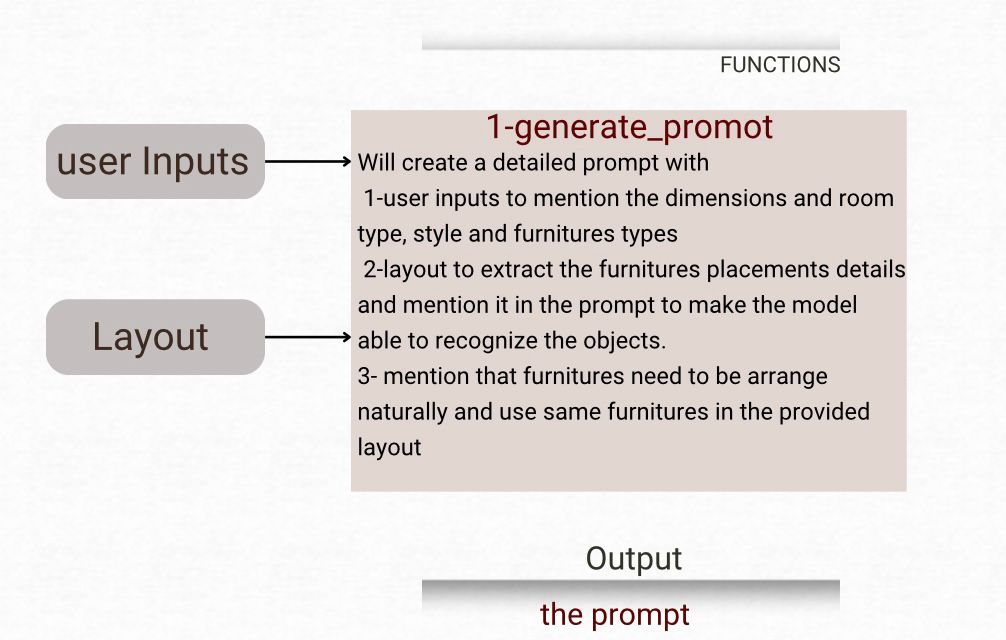}
    \caption{Generate the prompt.}
    \label{fig:fig3_generate_prompt}
\end{figure}

\begin{figure}[H]
    \centering
    \includegraphics[width=0.3\textwidth]{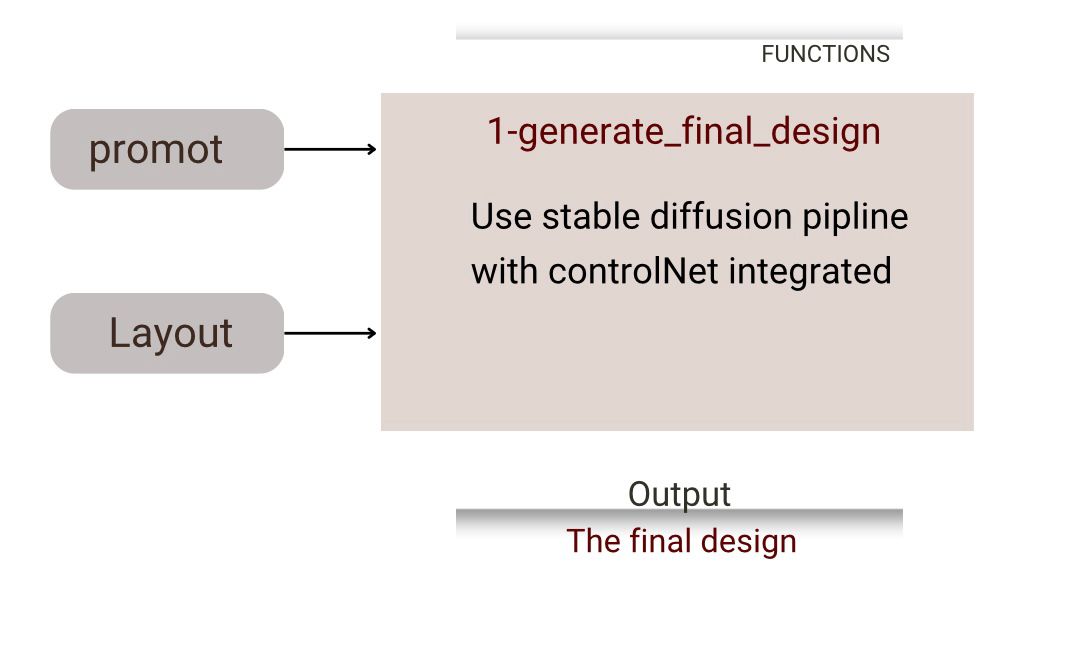}
    \caption{Generate the final design.}
    \label{fig:fig4_generate_design}
\end{figure}

   \item \textbf{Fine-tuning Classifiers:}
To ensure better results, we fine-tuned two classifiers based on the VGG16 architecture. One classifier was fine-tuned on room types datasets, and the other on room styles datasets. These classifiers are used to score the match between the user inputs (room type and style) and the generated design. If the design matches both the room type and style, the score will be high. However, if one or both do not match, the score will decrease.

-VGG16 Architecture: The VGG16 model is a deep convolutional network with 16 layers. It is used for image classification tasks, where it maps input images to predefined classes (room type and style, in our case) based on learned features.

\begin{figure}[H]
    \centering
    \includegraphics[width=0.3\textwidth]{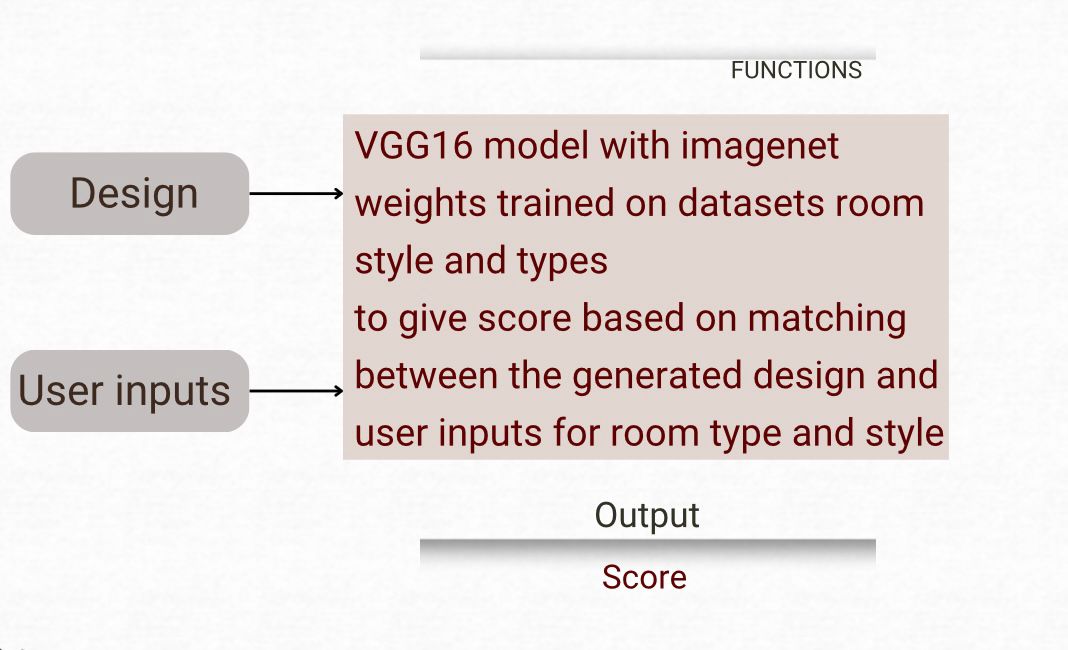}
    \caption{Evaluation.}
    \label{fig:fig5_evaluation}
\end{figure}
\end{enumerate}

\section{Experiments}
 In this section, we first introduce the experimental setup in
 Subsection A. Then, we present the results of qualitative
 evaluation and quantitative evaluation in Subsections B.1 and
 B.2.

\subsubsection*{A.Experimental Setup:}
 The DecoMind system was built using the Google Colab en
vironment, utilizing a Tesla T4 GPU. Three different datasets
 were used in this project (more details are provided in the Data
 Description section). Several models were integrated to build
 the system (all implementation details and specific usages of
 each model are explained in the Methodology section).

\subsubsection*{B.Evaluation of Results( Qualitative \& Quantitative):}
\subsubsection*{B.1 Qualitative Evaluation:}The DecoMind system demonstrated its effectiveness
 through the quality of the generated design images. The
 outputs closely matched the user’s textual descriptions, with
 only minor differences, if any. Examples of the system’s results
 are presented below in Figures 7 to 9 to illustrate its ability
 to align visual generation with user preferences.

\begin{figure}[H]
    \centering
    \includegraphics[width=0.3\textwidth]{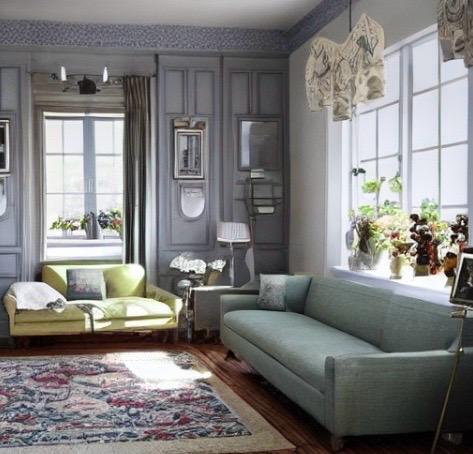}
    \caption{A generated modern living room design }
    \label{fig:labelname}
\end{figure}
\begin{figure}[H]
    \centering
    \includegraphics[width=0.3\textwidth]{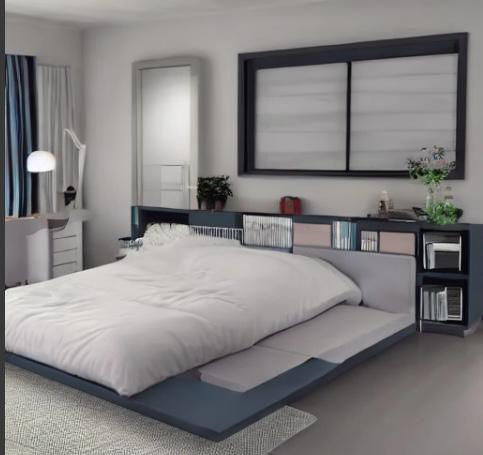}
    \caption{A generated modern bedroom design}
    \label{fig:labelname}
\end{figure}

\begin{figure}[H]
    \centering
    \includegraphics[width=0.3\textwidth]{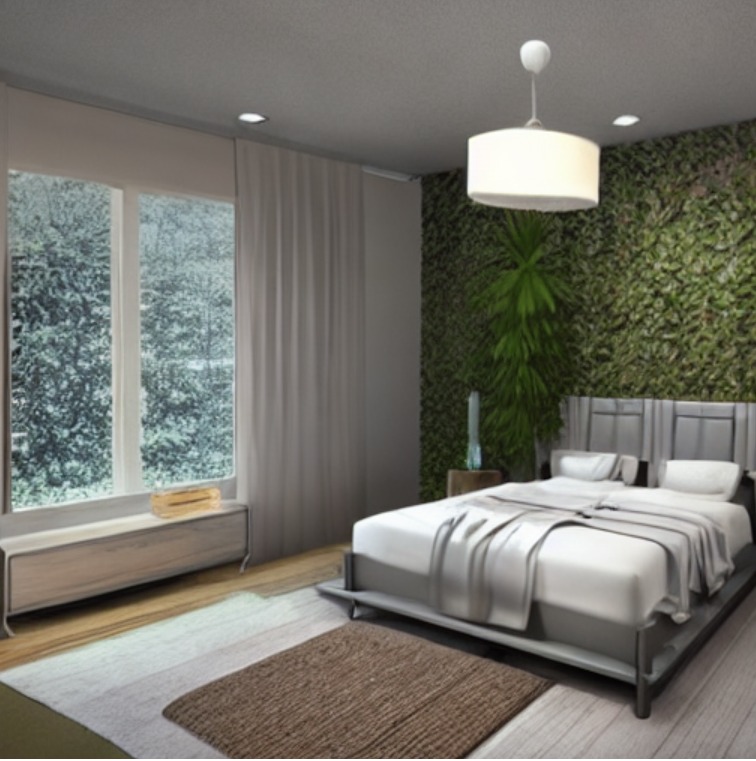}
    \caption{A generated minimalist bedroom design}
    \label{fig:labelname}
\end{figure}
\subsubsection*{B.2 Quantitative Evaluation:}
To quantitatively assess the performance of the DecoMind system, we employed two fine-tuned classifiers based on the VGG16 architecture to determine whether the generated designs aligned with the intended categories. Table 1 presents the evaluation results for five generated outputs. To conserve space, Only two generated designs (each showing a final design alongside its corresponding layout)  are presented in the paper.

\begin{figure}[!htbp]
    \centering
    % first image
    \begin{subfigure}[b]{0.32\textwidth}
        \centering
        \includegraphics[width=\linewidth]{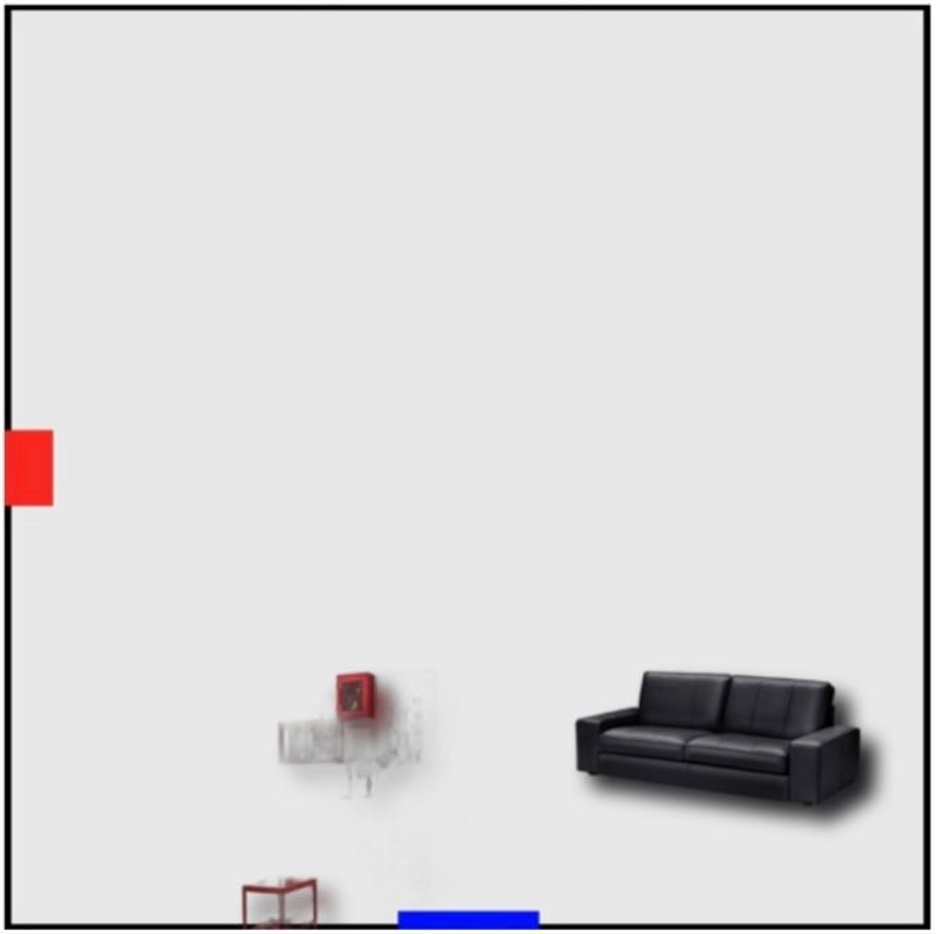}
        \caption{The generated layout}
        \label{fig:new2}
    \end{subfigure}\hfill
    % second image
    \begin{subfigure}[b]{0.32\textwidth}
        \centering
        \includegraphics[width=\linewidth]{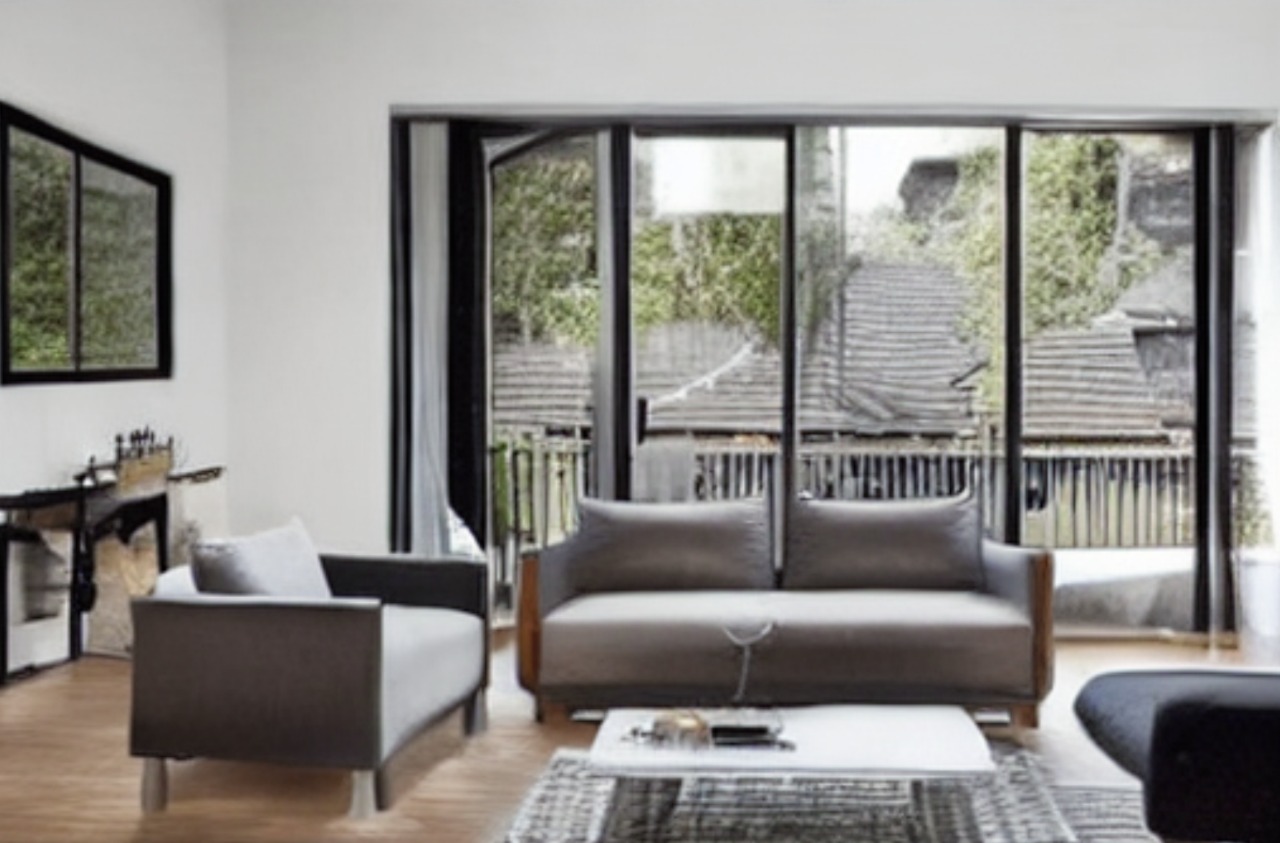}
        \caption{The generated image}
        \label{fig:fig10}
    \end{subfigure}\hfill
    % third image
    \begin{subfigure}[b]{0.32\textwidth}
        \centering
        \includegraphics[width=\linewidth]{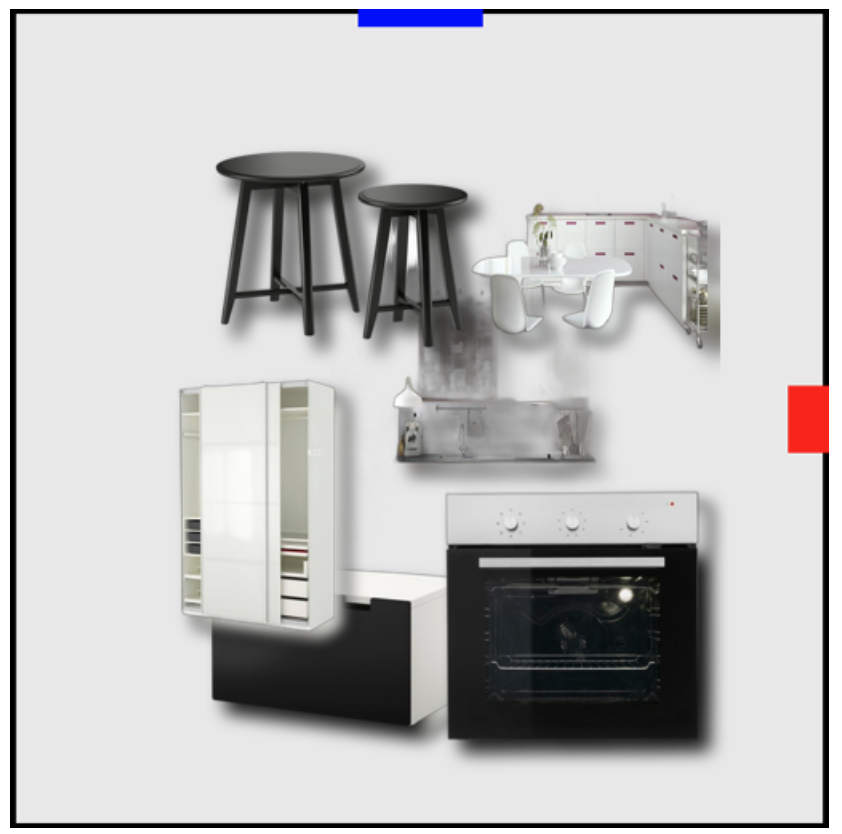}
        \caption{The generated layout}
        \label{fig:layout11}
    \end{subfigure}
    \caption{Generated layouts and images.}
    \label{fig:combined}
\end{figure}

\begin{figure}[H]
    \centering
    \includegraphics[width=0.32\textwidth]{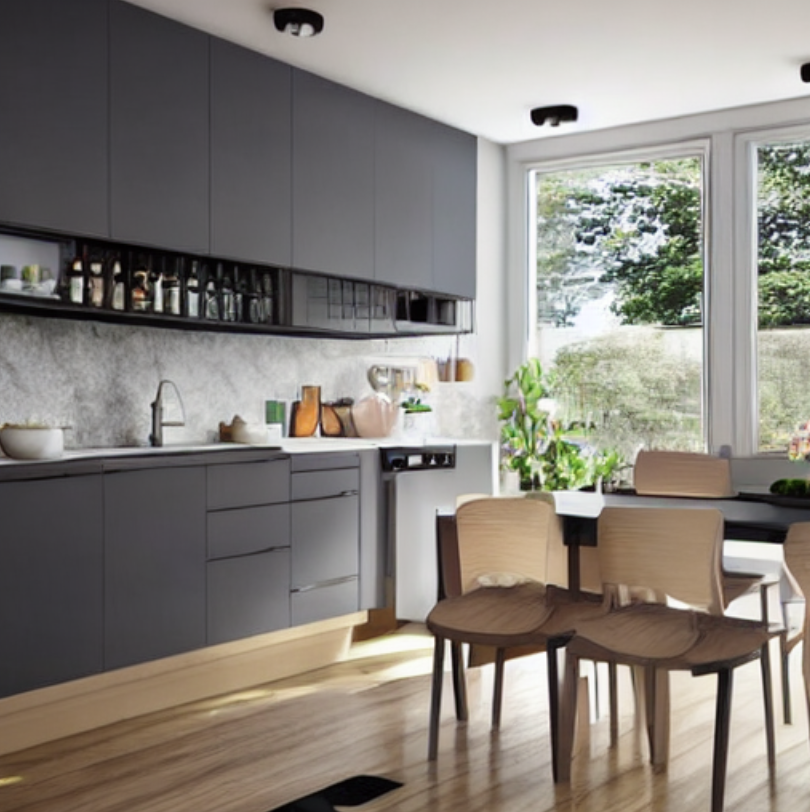}
    \caption{The generated image}
    \label{fig:labelname}
\end{figure}

\begin{table}[h]
    \centering
    \caption{Quantitative Evaluation Results for Multiple Generated Room Designs using Room Type and Style Classifiers}
    \label{tab:evaluation-results}
    \begin{tabular}{lccc}
        \toprule
        \textbf{Design ID} & \textbf{Room Type Match} & \textbf{Style Match} & \textbf{Final Score} \\
        \midrule
        Design 1 & Yes & No  & 0.50 \\
        Design 2 & Yes & No  & 0.50 \\
        Design 3 & Yes & No  & 0.50 \\
        Design 4 & No  & No  & 0.00 \\
        Design 5 & No  & No  & 0.00 \\
        \bottomrule
    \end{tabular}
\end{table}

\textbf{Analysis:} \\
The table above presents the evaluation of classifier outputs across five test cases. Three of the cases achieved partial success with correct room type but incorrect style, resulting in a final score of 0.5 each. The remaining two cases failed to match both room type and style, scoring 0.0.

\vspace{0.5em}

\textbf{Key Insight:} \\
A final score of 0.0 occurs when both the room type and the style do not match the expected values. This indicates a complete misclassification by the classifiers, which may be attributed to limitations in model accuracy, insufficient training quality, or data noise, despite the generated image clearly reflecting the intended room type and style.

\section{Discussion}
This section discusses the main limitations of the DecoMind system and the experiments conducted to address these issues followed by potential future enhancements.

\textbf{A. Main Limitations of the DecoMind System and Related Experiments:}

The Stable Diffusion model showed limitations in generating accurate furniture matches and realistic room layouts. We initially attempted to address this issue using the Inpainting variant, which preserved furniture consistency. However, the resulting images lacked realism and suffered from low visual quality, highlighting the need for further refinement.

To support the generation process and improve visual quality, enhancements were also made to the IKEA dataset. This included background removal to produce transparent images of individual furniture items, and filtering to exclude full-room images, retaining only standalone pieces. These enhancements significantly improved the model’s ability to interpret and integrate furniture elements into the generated layouts, resulting in more accurate outputs and better alignment with the intended room types and styles.

\begin{figure}[H]
        \centering
        \includegraphics[width=0.3\textwidth]{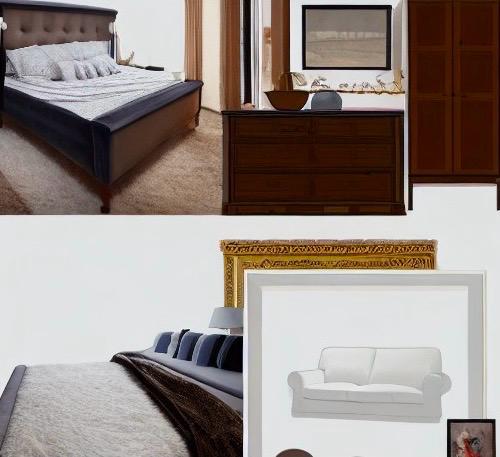} % Replace with your actual file name
        \caption{Result generated by the Inpainting Stable Diffusion model.Furniture is consistent, but the design lacks realism.}
        \label{fig:inpaint_result}
    \end{figure}

During experimentation with Stable Diffusion, the model occasionally attempted to generate two images of the same scene from different viewpoints to provide a multi-angle visualization of the design. However, This often failed either producing entirely different rooms or generating images from the same perspective with illogical or inconsistent additions.Figure 16 presents several examples of this issue.

\begin{figure}[H]
        \centering
        \includegraphics[width=0.3\textwidth]{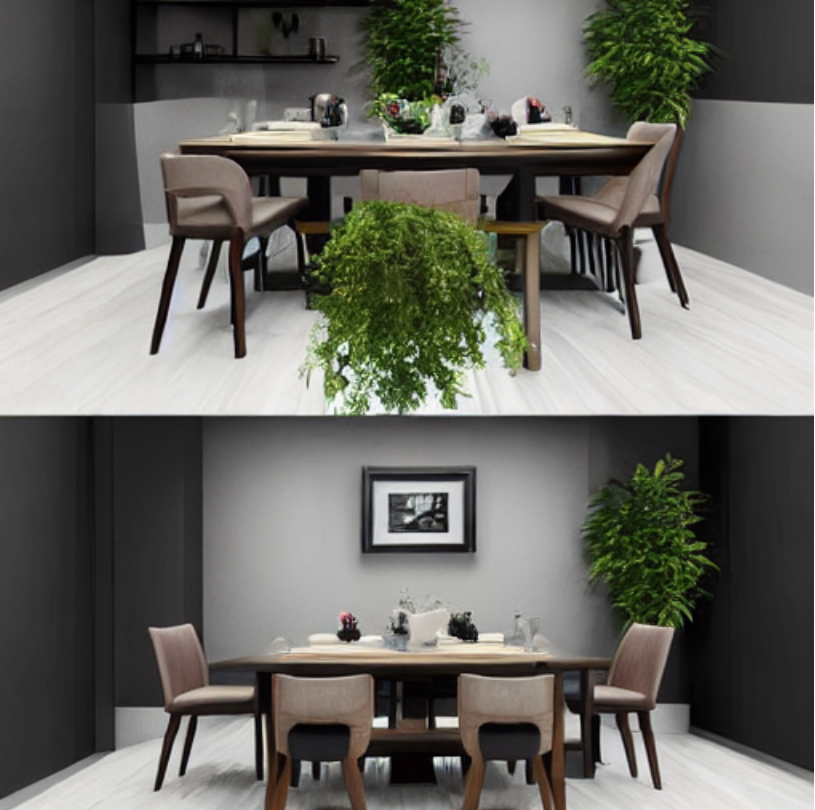} 
        \caption{Example of Inconsistent Multi-View Generation}
        \label{fig:inpaint_result}
    \end{figure}

\begin{figure}[H]
        \centering
        \includegraphics[width=0.3\textwidth]{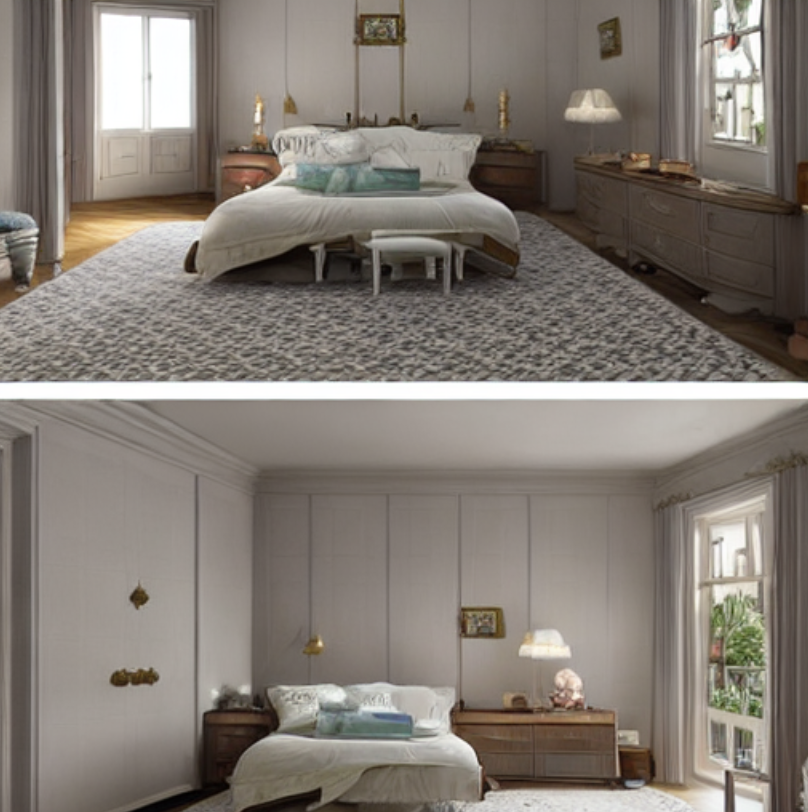} 
        \caption{Example of Inconsistent Multi-View Generation}
        \label{fig:inpaint_result}
    \end{figure}    

To enhance spatial realism, we explored 3D simulation techniques by adjusting layouts and crafting prompts that emphasized perspective (e.g., “a view from the room corner,” “showing both side walls”). We also attempted multi-view generation from different angles to mimic a 3D experience. While these strategies improved some visual cues, the model still lacked consistent depth and spatial coherence. This highlighted the core limitation: Stable Diffusion was not trained on 3D interior data. Fine-tuning on a 3D-specific dataset would likely resolve this, but was beyond our project’s scope due to time and resource constraints.

\textbf{B. Future Enhancements:}
\begin{itemize}
    \item Dataset Improvements:Integrate datasets from multiple furniture stores and include 3D views of furniture to provide more accurate and diverse furniture options, thus improving model generalization.
    \item CLIP Model Improvements: The furniture extraction and description capabilities of the CLIP model require fine-tuning to improve the accuracy of matching furniture types with user requests.
    \item  User Interaction Enhancements :
    
1- Custom Furniture Arrangement: Allow users to manually arrange furniture to provide greater flexibility and customization within the generated room layout.  

2-Feedback System: Implement an easy-to-use feedback mechanism that enables users to rate and suggest improvements to the generated designs, thereby supporting iterative system enhancement.
\end{itemize}

\section{Conclusion}
 In conclusion, this paper demonstrates the potential of
 combining CLIP, Stable Diffusion, and ControlNet to automate
 the process of generating interior design layouts based on
 user specifications. By extracting furniture, generating layout
 designs, and evaluating the results against room type and style
 classifiers,  individuals redesigning a single room or validating design choices before purchase, complementing rather than replacing professional interior designers.Despite the promising results, challenges remain in ensuring that the generated designs
 closely match the user’s selected furniture, and improving
 the performance of the CLIP model in extracting the most
 relevant furniture pieces. Nonetheless, the system represents a
 significant step forward in applying AI to the creative field of
 interior design, with future improvements aimed at increasing
 the realism and accuracy of the designs.

\appendix

\subsubsection*{A.Code Repository:}
The complete implementation of the DecoMind system, is available at the following public Google Colab link:

\begin{itemize}
    \item \href{https://colab.research.google.com/drive/1BM1swN_MggQTVOw_xTvXG1awVdZtdkN9?usp=sharing}{DecoMind Code(Click Here)}
\end{itemize}

\makebox[\columnwidth][l]{\bfseries B. More Generated Designs from the DecoMind System:}

\begin{figure}[H]
    \begin{minipage}{\textwidth}
        \vspace{0.2cm}
        \raggedright
       
        % First row: Bedroom + Dining
        \begin{minipage}{0.48\linewidth}
            \centering
            \textbf{Bedroom Designs}\par
            \begin{subfigure}[b]{0.48\linewidth}
                \includegraphics[width=\linewidth,height=4cm]{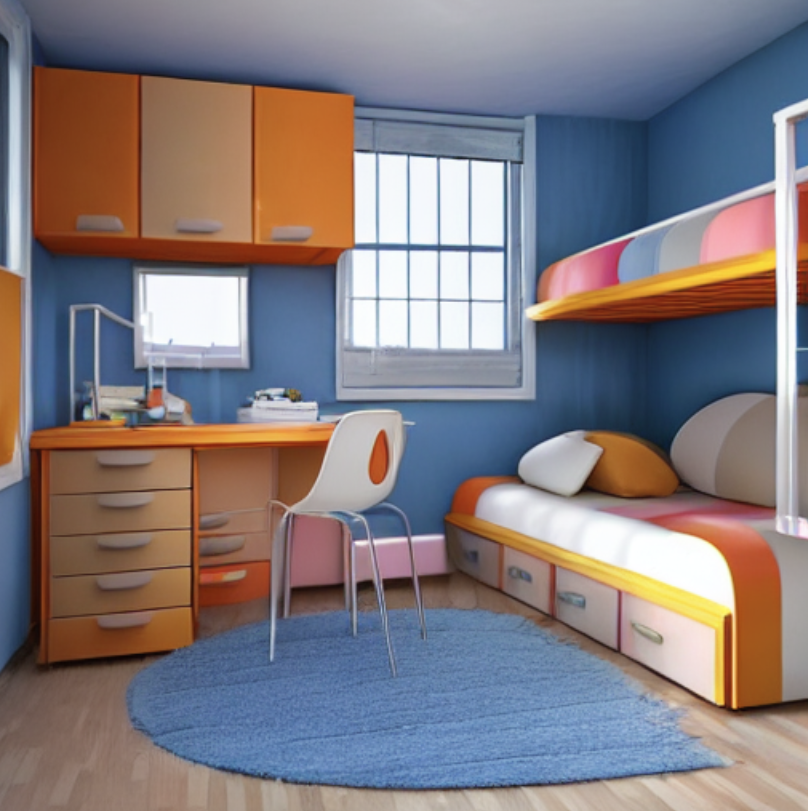}
                \caption{Bedroom 1}
            \end{subfigure}
            \begin{subfigure}[b]{0.48\linewidth}
                \includegraphics[width=\linewidth,height=4cm]{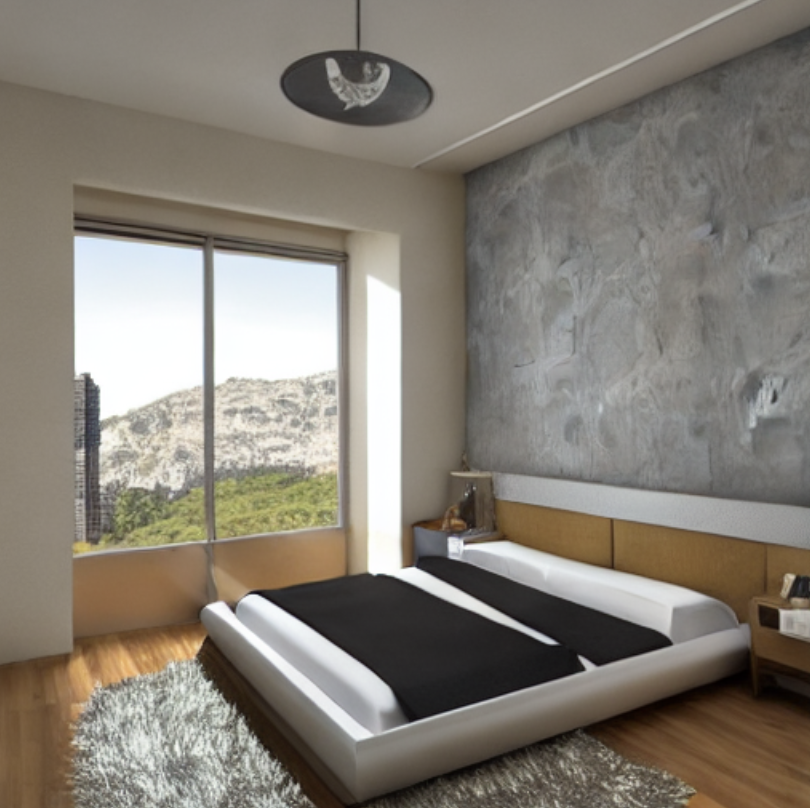}
                \caption{Bedroom 2}
            \end{subfigure}
            \begin{subfigure}[b]{0.48\linewidth}
                \includegraphics[width=\linewidth,height=4cm]{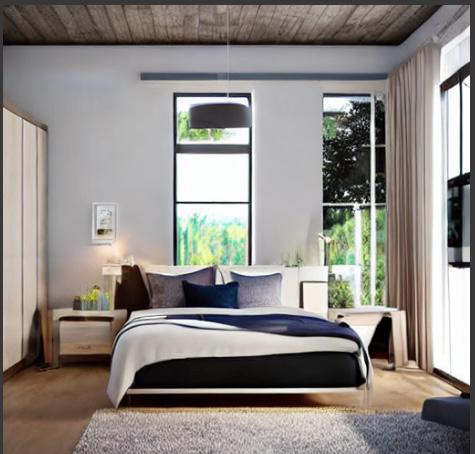}
                \caption{Bedroom 3}
            \end{subfigure}
            \begin{subfigure}[b]{0.48\linewidth}
                \includegraphics[width=\linewidth,height=4cm]{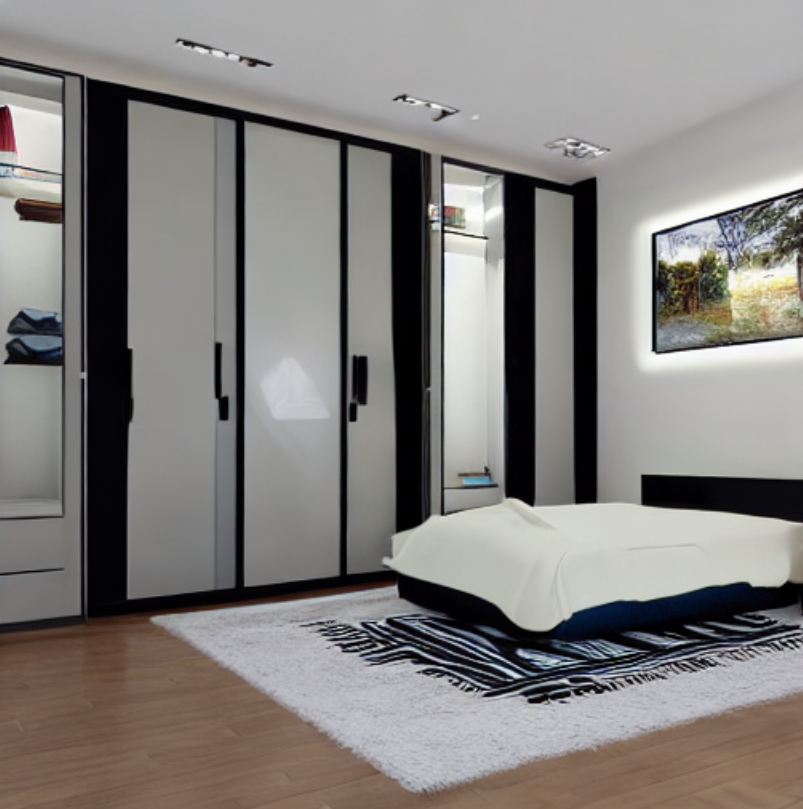}
                \caption{Bedroom 4}
            \end{subfigure}
        \end{minipage}
        \hfill
        \begin{minipage}{0.48\linewidth}
            \centering
            \textbf{Dining Room Designs}\par
            \begin{subfigure}[b]{0.48\linewidth}
                \includegraphics[width=\linewidth,height=4cm]{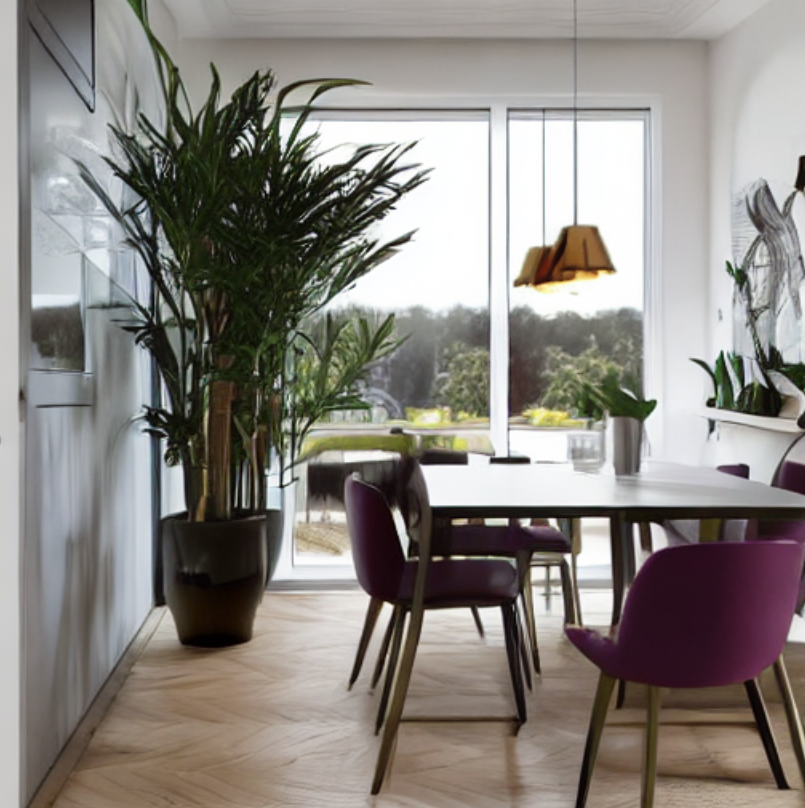}
                \caption{Dining Room 1}
            \end{subfigure}
            \begin{subfigure}[b]{0.48\linewidth}
                \includegraphics[width=\linewidth,height=4cm]{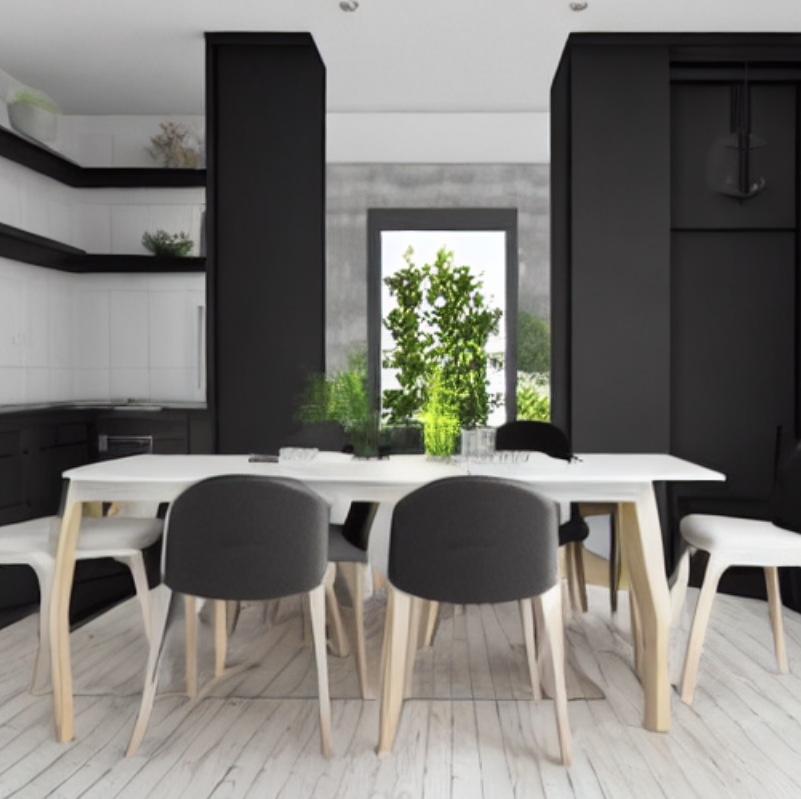}
                \caption{Dining Room 2}
            \end{subfigure}
            \begin{subfigure}[b]{0.48\linewidth}
                \includegraphics[width=\linewidth,height=4cm]{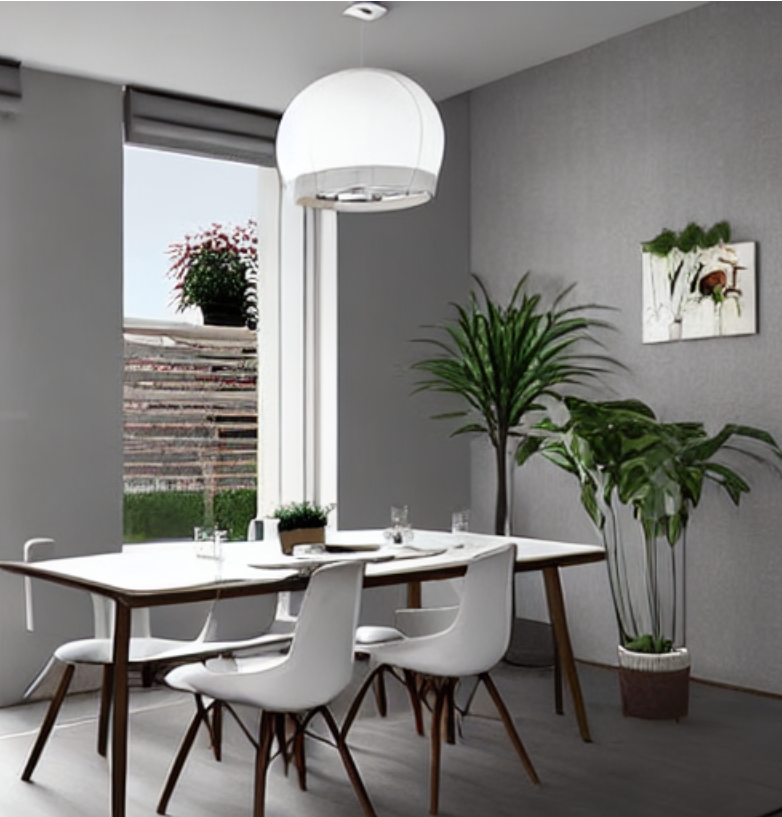}
                \caption{Dining Room 3}
            \end{subfigure}
            \begin{subfigure}[b]{0.48\linewidth}
                \includegraphics[width=\linewidth,height=4cm]{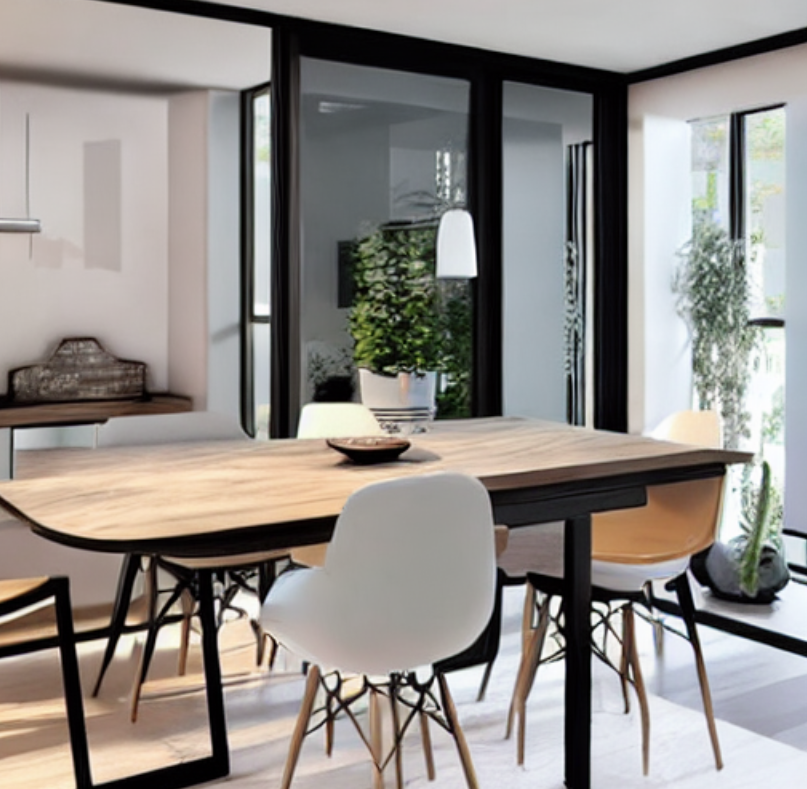}
                \caption{Dining Room 4}
            \end{subfigure}
        \end{minipage}

        \vspace{0.3cm}

        % Second row: Kitchen + Living
        \begin{minipage}{0.48\linewidth}
            \centering
            \textbf{Kitchen Designs}\par
            \begin{subfigure}[b]{0.48\linewidth}
                \includegraphics[width=\linewidth,height=4cm]{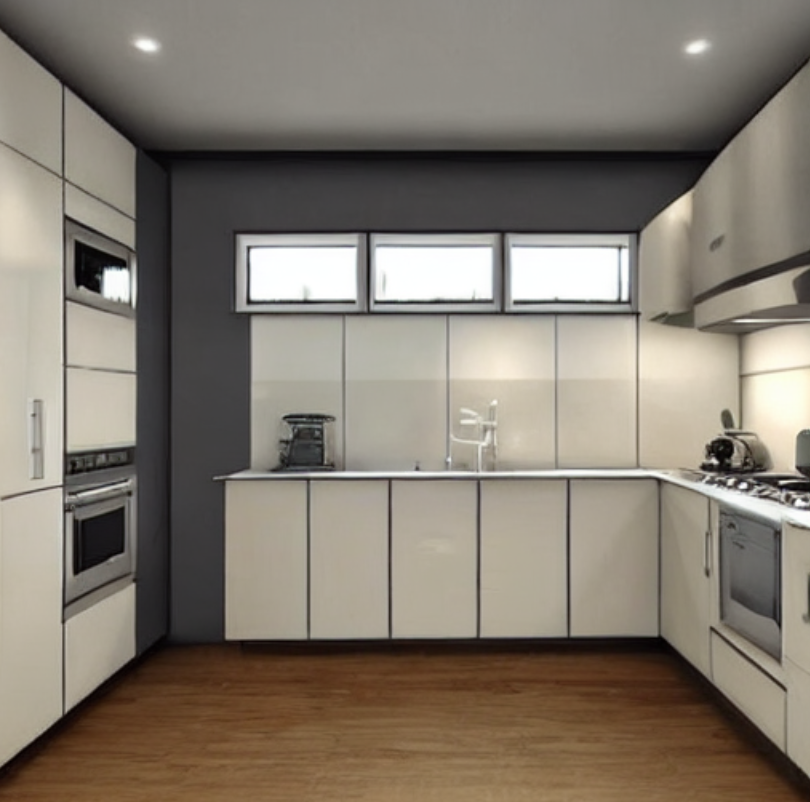}
                \caption{Kitchen 1}
            \end{subfigure}
            \begin{subfigure}[b]{0.48\linewidth}
                \includegraphics[width=\linewidth,height=4cm]{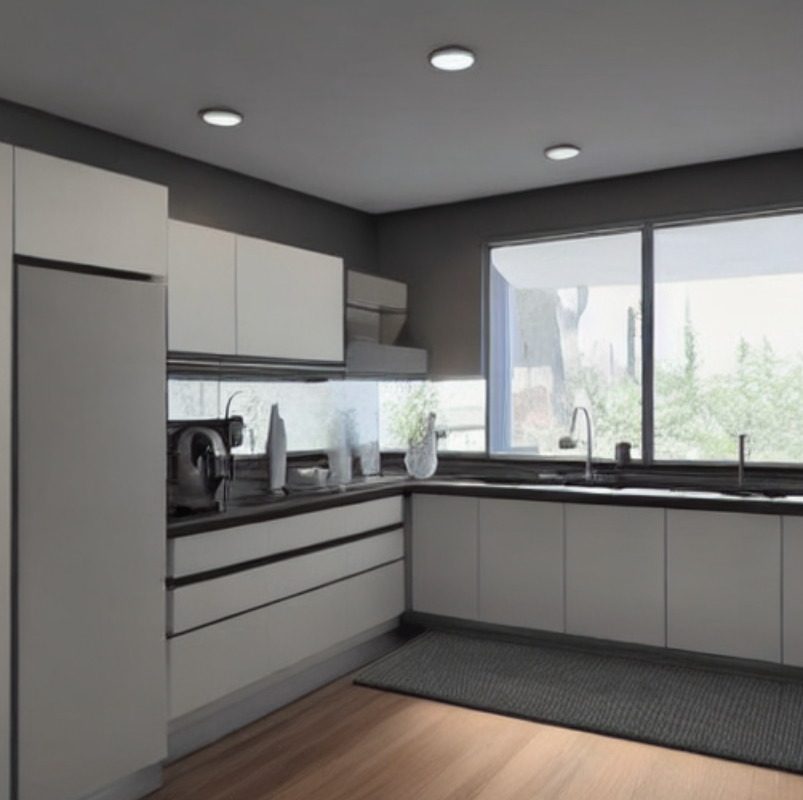}
                \caption{Kitchen 2}
            \end{subfigure}
            \begin{subfigure}[b]{0.48\linewidth}
                \includegraphics[width=\linewidth,height=4cm]{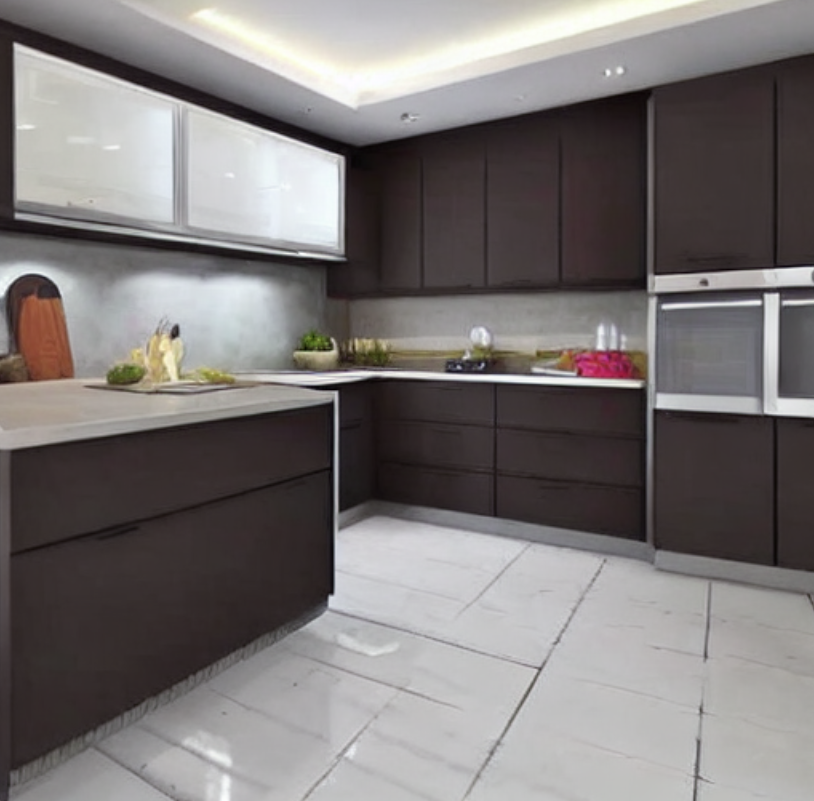}
                \caption{Kitchen 3}
            \end{subfigure}
            \begin{subfigure}[b]{0.48\linewidth}
                \includegraphics[width=\linewidth,height=4cm]{kitchen4.png}
                \caption{Kitchen 4}
            \end{subfigure}
        \end{minipage}
        \hfill
        \begin{minipage}{0.48\linewidth}
            \centering
            \textbf{Living Room Designs}\par
            \begin{subfigure}[b]{0.48\linewidth}
                \includegraphics[width=\linewidth,height=4cm]{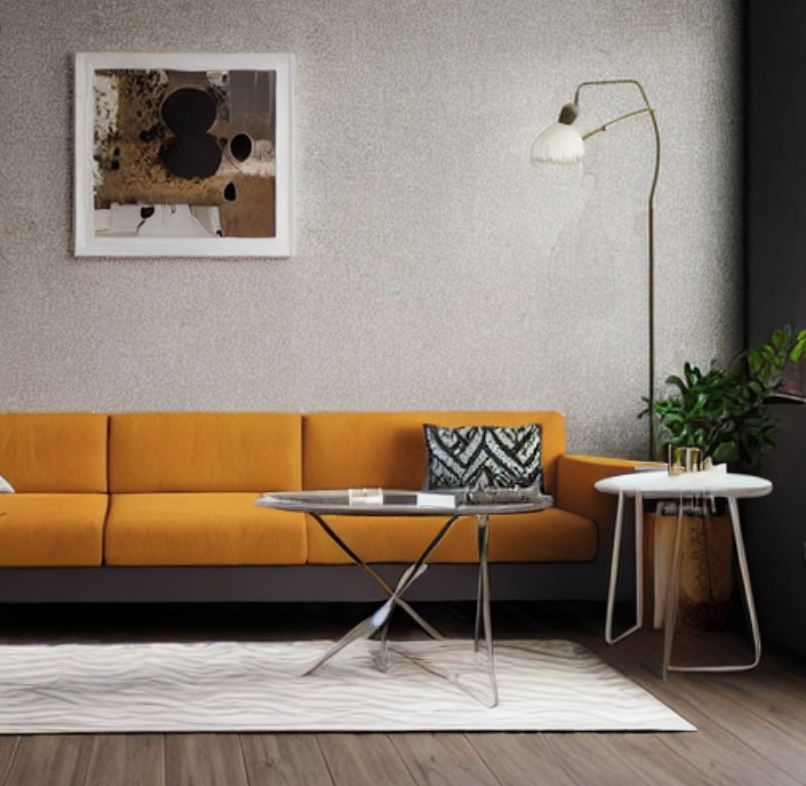}
                \caption{Living Room 1}
            \end{subfigure}
            \begin{subfigure}[b]{0.48\linewidth}
                \includegraphics[width=\linewidth,height=4cm]{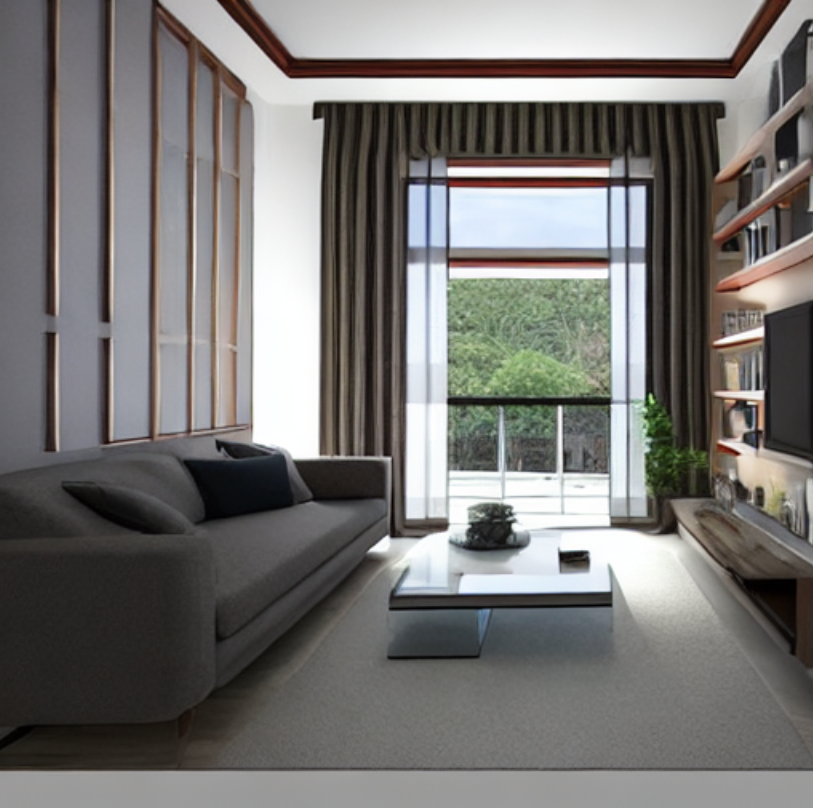}
                \caption{Living Room 2}
            \end{subfigure}
            \begin{subfigure}[b]{0.48\linewidth}
                \includegraphics[width=\linewidth,height=4cm]{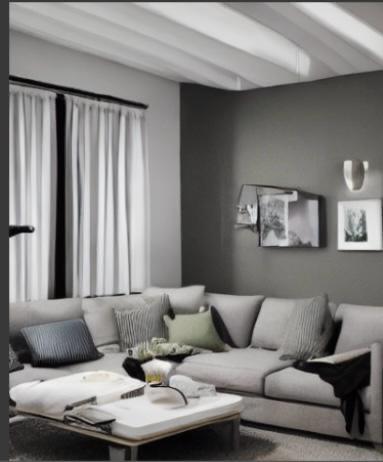}
                \caption{Living Room 3}
            \end{subfigure}
            \begin{subfigure}[b]{0.48\linewidth}
                \includegraphics[width=\linewidth,height=4cm]{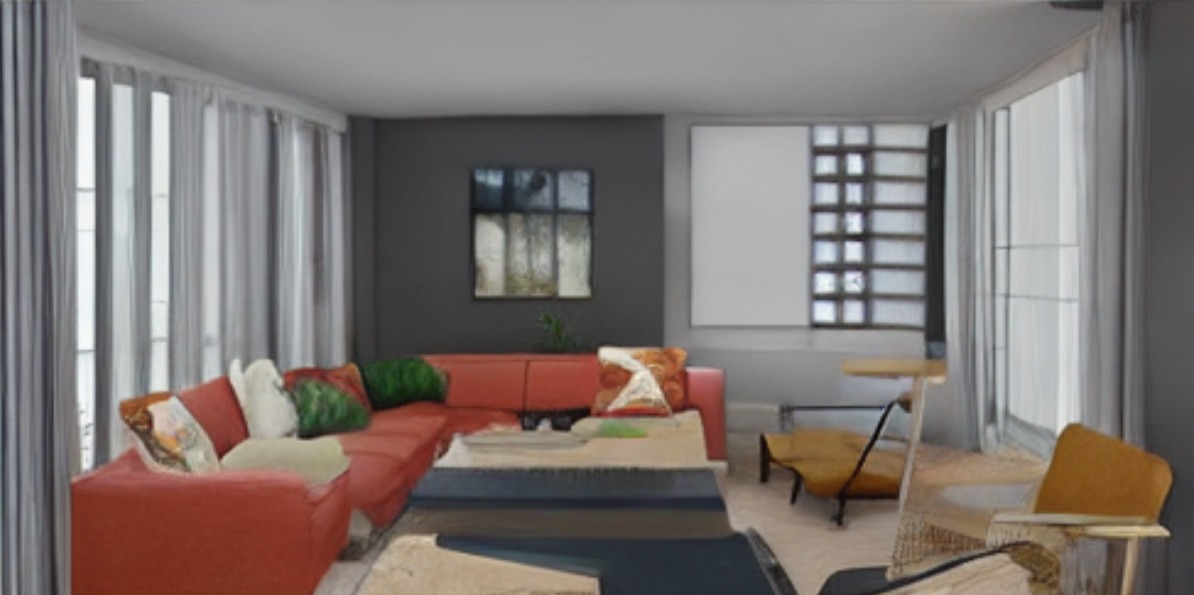}
                \caption{Living Room 4}
            \end{subfigure}
        \end{minipage}
        \vspace{0.3cm}

    \end{minipage}
    \captionsetup{justification=centering}
    \caption{Generated designs (Bedroom, Kitchen, Dining, Living) using the DecoMind system.}
    \label{fig:all_examples}
\end{figure}

\clearpage

\subsubsection*{C.References:} 
\renewcommand{\refname}{} 
\bibliographystyle{IEEEtran}
\bibliography{references}

\begin{thebibliography}{1}
\providecommand{\url}[1]{#1}
\csname url@samestyle\endcsname
\providecommand{\newblock}{\relax}
\providecommand{\bibinfo}[2]{#2}
\providecommand{\BIBentrySTDinterwordspacing}{\spaceskip=0pt\relax}
\providecommand{\BIBentryALTinterwordstretchfactor}{4}
\providecommand{\BIBentryALTinterwordspacing}{\spaceskip=\fontdimen2\font plus
\BIBentryALTinterwordstretchfactor\fontdimen3\font minus \fontdimen4\font\relax}
\providecommand{\BIBforeignlanguage}[2]{{%
\expandafter\ifx\csname l@#1\endcsname\relax
\typeout{** WARNING: IEEEtran.bst: No hyphenation pattern has been}%
\typeout{** loaded for the language `#1'. Using the pattern for}%
\typeout{** the default language instead.}%
\else
\language=\csname l@#1\endcsname
\fi
#2}}
\providecommand{\BIBdecl}{\relax}
\BIBdecl

\bibitem{cliplayout2023}
J.~Liu, W.~Xiong, I.~Jones, Y.~Nie, A.~Gupta, and B.~Oğuz, ``Clip-layout: Style-consistent indoor scene synthesis with semantic furniture embedding,'' \emph{arXiv preprint arXiv:2303.03565}, March 2023.

\bibitem{deepfurniture2020}
B.~Liu, J.~Zhang, X.~Zhang, W.~Zhang, C.~Yu, and Y.~Zhou, ``Furnishing your room by what you see: An end-to-end furniture set retrieval framework with rich annotated benchmark dataset,'' \emph{arXiv preprint arXiv:1911.09299}, January 2020.

\bibitem{creativediffusion2025}
Z.~Xu, Y.~Zhang, H.~Luo, and Y.~Wang, ``Creative interior design matching the indoor structure generated through diffusion model with an improved control network,'' \emph{Frontiers of Architectural Research}, vol.~14, no.~3, pp. 614--629, 2025.

\bibitem{ikeafurniture2025}
P.~AI, ``Ikea furnitures computer vision dataset,'' \url{https://universe.roboflow.com/projet-ai/ikea-furnitures/dataset/2/download}, accessed: May 04, 2025.

\bibitem{houserooms2025}
R.~Reni, ``House rooms image dataset,'' \url{https://www.kaggle.com/datasets/robinreni/house-rooms-image-dataset}, accessed: May 04, 2025.

\bibitem{interiordesignstyles2025}
S.~Arullin, ``Interior design styles,'' \url{https://www.kaggle.com/datasets/stepanyarullin/interior-design-styles?utm_source=chatgpt.com}, accessed: May 04, 2025.

\end{thebibliography}

\end{document}